\documentclass[10pt,conference]{IEEEtran}
\IEEEoverridecommandlockouts
\usepackage{cite}
\usepackage{hyperref}
\usepackage{mathtools}
\usepackage{amsmath,amssymb,amsfonts}
\usepackage{graphicx}
\usepackage{textcomp}
\usepackage{soul}
\usepackage[table]{xcolor}
\usepackage{etc}
\usepackage{algorithm}
\usepackage{algpseudocode}
\usepackage{booktabs}
\usepackage[normalem]{ulem}
\usepackage{subcaption}
\usepackage{enumitem}
\usepackage{balance}

\usepackage{multirow,array}

\newcolumntype{C}[1]{>{\centering\arraybackslash}p{#1}}

\definecolor{darkgreen}{rgb}{0.0, 0.5, 0.0}

\def\BibTeX{{\rm B\kern-.05em{\sc i\kern-.025em b}\kern-.08em
    T\kern-.1667em\lower.7ex\hbox{E}\kern-.125emX}}

\definecolor{codegreen}{rgb}{0.1,0.7,0.1}
\definecolor{codegray}{rgb}{0.5,0.5,0.5}

\newtcolorbox{rqbox}[1][]{
    colback=green!10,
    colframe=codegreen,
    arc=1mm,
    boxrule=0.5pt,
    coltitle=black,
    fonttitle=\bfseries,
    title=#1
}

\newtcolorbox{rqboxgray}[1][]{
    colback=codegray!10,
    colframe=codegray,
    arc=1mm,
    boxrule=0.5pt,
    coltitle=black,
    fonttitle=\bfseries,
    title=#1
}

\definecolor{green}{rgb}{0.0, 0.5, 0.0}

\newcommand*\circled[1]{\tikz[baseline=(char.base)]{
            \node[shape=circle,draw,fill=black,text=white,inner sep=1pt] (char) {#1};}}

\definecolor{errorred}{RGB}{220, 50, 47}

\lstdefinestyle{github}{
    backgroundcolor=\color[HTML]{F8F9FA},
    basicstyle=\footnotesize\ttfamily\color[HTML]{333333},
    commentstyle=\color[HTML]{EF6565},
    keywordstyle=\color[HTML]{3498DB}\bfseries,
    numberstyle=\tiny\color[HTML]{BDC3C7},
    stringstyle=\color[HTML]{446EBB},
    identifierstyle=\color[HTML]{3E3F3D},
    numbers=left,
    numbersep=10pt,
    frame=leftline,
    framerule=0.8pt,
    framesep=8pt,
    framextopmargin=6pt,
    framexbottommargin=6pt,
    xleftmargin=10pt,
    rulecolor=\color{gray!30},
    tabsize=4,
    breaklines=true,
    captionpos=b,
    showspaces=false,
    showstringspaces=false
}

\begin{document}

\title{
Watts This Smell: A Comprehensive Taxonomy of Software Energy Smells
}

\author{
\IEEEauthorblockN{Mohammadjavad Mehditabar, Saurabhsingh Rajput, Tushar Sharma}
\IEEEauthorblockA{\textit{Dalhousie University, Halifax, Canada} \\
\textit{\{javad, saurabh, tushar\}@dal.ca}}
}

\maketitle

\begin{abstract}
As software proliferates across domains, its aggregate energy footprint has become a major concern.
To reduce software's growing environmental footprint, developers need to identify and refactor \textit{energy smells}: source code implementations, design choices, or programming practices that lead to inefficient use of computing resources.
Existing catalogs of such smells are either domain-specific, limited to performance anti-patterns, lack fine-grained root cause classification, 
or remain unvalidated against measured energy data.
In this paper, we present a comprehensive, language-agnostic, taxonomy of software energy smells. Through a systematic literature review of 60 papers and exhaustive snowballing, we coded 320 inefficiency patterns into 12 primary energy smells and 65 root causes mapped to the primary smells. 
To empirically validate this taxonomy, we profile over 21,000 functionally equivalent Python code pairs for energy, time, and memory, and classified the top 3000 pairs by energy difference using a multi-step LLM pipeline, mapping 55 of the 65 root causes to real code. The analysis reveals that 71\% of samples exhibit multiple co-occurring smells, memory-related smells yield the highest per-fix energy savings, while power draw variation across patterns confirms that energy optimization cannot be reduced to performance optimization alone. 
Along with the taxonomy, we release the labeled dataset, including energy profiles and reasoning traces, to the community. 
Together, they provide a shared vocabulary, actionable refactoring guidelines, and an empirical foundation for energy smell detection, energy-efficient code generation, and green software engineering at large.
\end{abstract}


\section{Introduction}
\label{sec:intro}
Kent Beck defined code smells as code structures that suggest, and sometimes scream for, the possibility of refactoring~\cite{fowler2018refactoring}.
The existence of these smells reduces code quality and negatively impacts maintainability of a software system~\cite{garcia2009toward,khomh2011bdtex}.
Code smells are classified into different categories, types, and variants based on their occurrence granularity, artifact, and scope~\cite{sharma2018survey}. It includes smells in traditional code, such as implementation ~\cite{fowler2018refactoring,brown1998antipatterns}, design ~\cite{binkley2008dependence,suryanarayana2014refactoring}, architecture smells ~\cite{garcia2009toward,Sharma2020}, test smells~\cite{van2001refactoring,greiler2013automated}, and configuration smells ~\cite{sharma2016does}. 

Similar to traditional code smells that mainly target the maintainability attribute of software quality, smells also appear in code that affect other quality aspects, such as energy efficiency.
Energy smells are defined as source code anti-patterns that use hardware resources inefficiently, leading to unnecessary energy consumption at runtime~\cite{vetro2013definition,gottschalk2012removing}. By identifying such smells, analogous to how traditional code smells guide maintainability improvements, developers can refactor them to improve energy efficiency and overall software performance and quality.

A common misconception in software engineering is treating execution performance and energy consumption as interchangeable metrics. 
Although energy correlates with performance~\cite{pallister2015identifying,chan2020investigating}, the two measure distinctly different phenomena. 
Performance primarily reflects execution time, \ie{} how long the CPU actively processes instructions to complete a task~\cite{patterson2016computer}. 
Energy consumption, in contrast, depends on how code interacts with underlying hardware states, power domains, and dynamic power management systems~\cite{li2011energy,carroll2014unifying}.  
Critically, faster execution does not always lead to lower energy consumption~\cite{pereira2017energy,weber2023twins}.
Since energy equals power multiplied by time ($E = P \times T$), and power draw varies across code patterns~\cite{pereira2017energy}, a code change that reduces execution time can simultaneously increase power draw, resulting in higher total energy.
This effect is well-documented in concurrent programming, where parallelism trades time for increased resource utilization~\cite{abdulsalam2015using,zambre2016parallel,goetz2006java}. 

Consequently, systems tuned exclusively for speed can harbor energy inefficiencies invisible to conventional profilers.

Beyond this conceptual distinction, the global software ecosystem is rapidly expanding. This growth is fueled by cloud computing, mobile platforms, microservices, and IoT deployments, and is expected to continue due to the rise of AI-generated software~\cite{cui2026effects}. As software scales in size and deployment volume, its overall energy footprint increases accordingly. Even small implementation-level inefficiencies, when replicated across millions of running instances and billions of invocations, translate into significant energy waste~\cite{lannelongue2021green,andrae2015global}. 
While energy inefficiency ultimately manifests at the level of hardware micro-architecture and power management, it is driven by high-level software decisions: algorithmic choices, data structure selection, API usage patterns, and implementation idioms that developers make long before any hardware is involved~\cite{georgiou2017iot,ournani2020reducing,georgiou2019software}.
These factors together motivate the systematic study of energy smells in software engineering. Just as code smells equip the community with a shared vocabulary of structural anti-patterns signaling poor design, which subsequently operationalized into static analysis tools, refactoring catalogs, and quality metrics~\cite{fowler2018refactoring, sharma2018survey}, a well-defined taxonomy of energy smells can empower developers with actionable guidance for energy-aware implementation.

Despite existing catalogs of performance anti-patterns~\cite{smith2000software,smith2002new,smith2003more,dargan2025my,tao2024beyond,zhao2020performance} and energy anti-patterns~\cite{brandolese2002impact,gottschalk2012removing,vetro2013definition}, several critical gaps persist across this body of work. 
First, most performance-focused catalogs~\cite{smith2000software,smith2002new,smith2003more,dargan2025my,tao2024beyond,zhao2020performance} define their taxonomies in terms of time efficiency, omitting energy consumption as a distinct optimization target.
Second, existing taxonomies have a limited scope and do not comprehensively cover different types of smells.
Dargan~\etal{}~\cite{dargan2025my} and Tao~\etal{}~\cite{tao2024beyond}, derived from student submissions, cover narrow problem domains and omit implementation-level concerns such as concurrent programming and hardware locality behavior.
Third, several catalogs are domain- or language-specific by design, limiting their applicability. For example, the catalog proposed by Gottschalk~\etal{}~\cite{gottschalk2012removing} covers Android-specific energy bugs derived from a literature survey.
Fourth, validation of the catalog or energy profiling is either missing in current studies~\cite{gottschalk2012removing,gurung2024static} or conducted within a narrowly scoped setup~\cite{vetro2013definition}.

To address these limitations, we build on existing energy and performance smell catalogs, identify a set of known issues leading to energy inefficiency, filter and consolidate them, and develop a comprehensive, language-agnostic, two-level taxonomy of software energy smells. The first level defines general energy smell categories, and the second level pinpoints the actionable and precise root causes. To ensure exhaustive coverage, we conduct a systematic literature review supplemented by forward and backward snowballing, encompassing $60$ articles across diverse domains that identify issues leading to performance or energy inefficiencies. Following a multi-phase filtering and annotation process, two independent annotators apply open coding, axial coding, and selective coding to identify $320$ relevant issues (Cohen's $\kappa = 0.74$). This process produces a consolidated \textbf{taxonomy of twelve energy smell categories and 65 root causes}.
To empirically validate this taxonomy, we construct a profiled Python dataset based on \texttt{Pie-Perf}~\cite{madaan2023learning}, containing pairs of functionally equivalent code snippets with measured energy, time, and memory consumption. We apply a multi-step classification pipeline using a state-of-the-art reasoning LLM (\texttt{DeepSeek-V3.2}~\cite{deepseekai2025deepseekv32}) to map the root causes of inefficiency in $3,000$ code pairs to the taxonomy, successfully identifying all twelve categories across $85\%$ of our identified root causes, validating the real-world prevalence and relevance of our taxonomy.

In summary, this paper makes the following primary contributions:
\begin{itemize}

\item \textbf{A comprehensive taxonomy} of twelve energy smell categories and 65 root causes, derived from a systematic literature review. The taxonomy is language-agnostic, covering a variety of smells from low-level hardware effects to high-level algorithmic choices.

\item \textbf{A labeled dataset} of $21,428$ code pairs with verified energy, memory, and time measurements, including a manually validated subset of $3,000$ pairs annotated with multi-label smell classifications and step-by-step reasoning traces to support future research.

\item \textbf{An empirical analysis} of the prevalence, co-occurrence, and energy impact of each smell category, providing evidence that energy optimization cannot be reduced to performance optimization alone.

\end{itemize}

All artifacts, including the taxonomy, dataset, and annotations, are publicly available in our replication package~\cite{anonymous_2026_18896365}.

\section{Methodology}
\label{sec:methodology}

\subsection{Taxonomy Design Goals}
\label{subsec:design-goals}
\textbf{Energy smells} \textit{are source code implementations, design choices, or programming practices that lead to inefficient use of computing resources}~\cite{vetro2013definition,gottschalk2012removing}.
The primary \textbf{objective} of this research is to define a comprehensive, well-detailed, and language-agnostic taxonomy of software energy smells. To ensure broad applicability across diverse programming paradigms and application domains, the taxonomy must systematically categorize inefficiencies ranging from implementation-level issues to high-level algorithmic flaws. To achieve this, we structure the taxonomy into a two-level hierarchy: 
\\
\textbf{Energy smell (Category)}: A high-level, observable issue in source code that uses computing resources inefficiently.
\\
\textbf{Root cause (Subcategory)}: The specific, actionable non-optimal or sub-optimal implementation choice---such as a poorly chosen data structure or a redundant loop---that triggers the broader energy waste.

A robust taxonomy requires that an inefficient implementation found in any application maps seamlessly to our defined categories and sub-categories. While strict mutual exclusivity is not a hard constraint---since programming logic often intertwines and one smell may correlate with or trigger another---we aim to design the taxonomy to minimize overlap. Each energy smell and its underlying root cause needs to maintain a clear boundary. This precise separation ensures that developers can pinpoint the exact source of an inefficiency, facilitating straightforward refactoring and classification.

\begin{figure*}[t]
    \centering
    \includegraphics[width=1\textwidth]{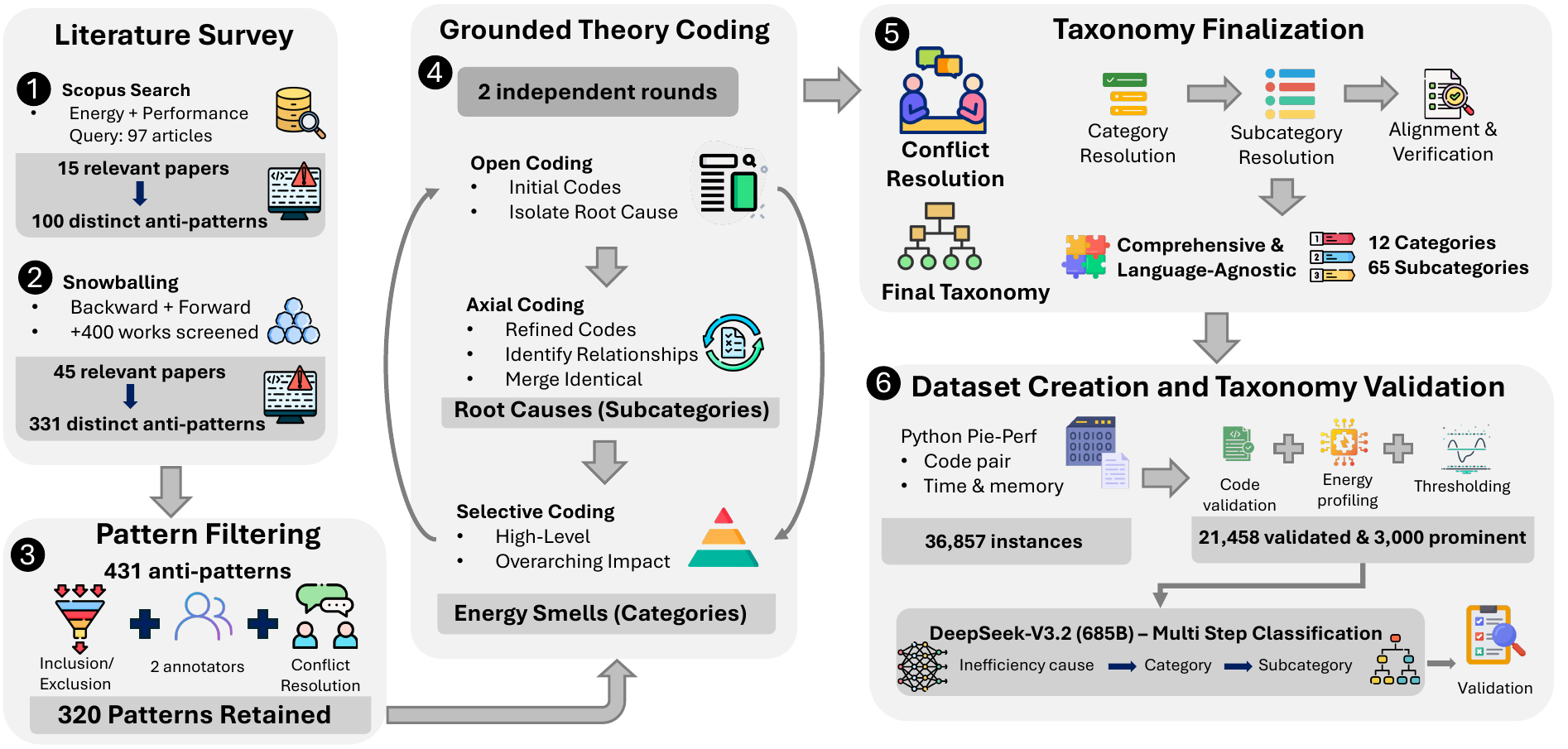}
    \caption{Overview of the methodology }
    \vspace{3mm}
    \label{fig:methodology}
\end{figure*}

To systematically capture and classify these inefficiencies,
we perform a rigorous, multi-phase literature review and qualitative coding process, following established software engineering systematic mapping guidelines~\cite{petersen2015guidelines}. Fig.~\ref{fig:methodology} illustrates the full pipeline.

\subsection{Phase 1: Primary Literature Search}
\label{subsec:scopus}
We initiate the literature search and collection process of relevant articles by querying Scopus, illustrated in the Step \circled{1} of Fig.~\ref{fig:methodology}. Scopus is a comprehensive database widely utilized in software engineering for literature reviews due to its extensive coverage and advanced search capabilities~\cite{poy2025impact,garousi2016citations}. We construct a targeted query designed to identify articles discussing code-level efficiency anti-patterns:
\textit{TITLE-ABS-KEY ((``energy'' OR ``power consumption'' OR ``green'' OR ``performance inefficiency'') AND (``code smell'' OR ``smell'' OR ``anti-pattern'') AND (``software'' OR ``code'')) AND (LIMIT-TO (SUBJAREA, ``COMP''))}

We explicitly include ``performance inefficiency'' because software performance and energy consumption are highly correlated metrics~\cite{pallister2015identifying,chan2020investigating}. While strictly defined energy smells often relate to how code mismanages physical hardware components (\eg{} preventing CPU sleep states or triggering excessive memory fetches)~\cite{li2011energy,carroll2014unifying}, total energy consumption is fundamentally governed by execution time as expressed by $E=P \times T$. Also, the literature detailing performance anti-patterns can contribute significantly to our taxonomy, because these issues are often less application-specific~\cite{dargan2025my,zhao2020performance} than pure energy studies~\cite{gottschalk2012removing,vetro2013definition}. Hence, including these issues will help meet our goal of broader applicability. Nonetheless, we evaluate performance findings cautiously to ensure they legitimately impact energy consumption.

The initial Scopus search yields $97$ relevant articles. We apply strict inclusion and exclusion criteria to filter this set. 
We \textit{include} any study that identifies, defines, or empirically proves a transferable performance or energy inefficiency at the code level, regardless of its original domain (\eg{} extracting language-agnostic practices from infrastructure-specific studies~\cite{kosbar2025smells}). We also include studies discussing specific logic and implementation issues, from implementation-level to high algorithmic-level. 
We \textit{exclude} studies on design smells, \eg{} feature envy, that primarily target maintainability rather than energy efficiency directly. 
We also exclude duplicate reports, vague definitions, and overly niche, application-specific implementations that cannot transfer to other domains.
Notably, about 29\% of the captured articles focus on Android, 24\% examine general software inefficiencies or the energy impact of refactoring code smells and design choices, and 11\% target domains such as \texttt{web}, \texttt{cloud}, \texttt{IoT}, and \texttt{SQL}. The remainder address other software smells (e.g., test smells or code smell detection), with a small portion being irrelevant to our search. After this filtering, we retain $15$ relevant studies, extracting $100$
distinct performance and energy anti-patterns along with their definitions.

\subsection{Phase 2: Snowballing}
\label{subsec:snowballing}

To maximize coverage beyond the initial Scopus results, we use the 15 retained articles as seed papers and conduct comprehensive snowballing.
Then, we capture all code snippets in papers retrieved from snowballing reported as being energy- or performance-inefficient, as depicted in Step \circled{2} of Fig.~\ref{fig:methodology}.

We begin with backward snowballing on all papers identified through the Scopus search. For each newly discovered paper, we apply the same extraction procedure and recursively examine its references and related work sections. In addition, we perform forward snowballing on highly relevant and foundational studies. 
In particular, performance anti-pattern studies by Smith~\etal{}~\cite{smith2000software,smith2002new,smith2003more} are extensively expanded forward, as subsequent papers citing them frequently propose additional relevant issues.

Additionally, we include grey literature referenced within primary or snowballed sources when it reports validated inefficiencies. 
For example, we include the \texttt{CAST} Software's $979$ green rules, derived from accepted standards such as PCI~\cite{pci_ssc_dss}, MITRE's CWE~\cite{mitre_cwe}, and OMG's CISQ~\cite{omg_cisq}.
Overall, we screen more than $400$
works to identify studies reporting performance or energy anti-patterns. 

After filtering out studies that merely repeat previously identified issues and that comply with our inclusion and exclusion criteria, we retain a total of $45$ studies, among them $19$ targeting specifically energy-efficiency and $26$ performance.
Combined with the $15$ Scopus papers, this yields a final corpus of $60$ relevant studies. From these, we extract \textbf{331 distinct issues}, bringing the total identified potential energy-efficiency issues to $\textbf{431}$ ($100$ from Scopus + $331$ from snowballing).

\subsection{Phase 3: Relevance Filtering}
\label{subsec:filtering}
In Step~\circled{3} of Fig.~\ref{fig:methodology}, we classify each of the $431$ collected issues as \textit{relevant} or \textit{irrelevant}. The primary objective is to retain language-agnostic constructs that directly impact energy consumption, time complexity, or space complexity. To ensure an unbiased process, the first two authors, each with more than a year of experience in software energy consumption research, independently label all the extracted items based on the predefined inclusion and exclusion criteria.
Out of $431$ total items, $43$ result in disagreement between the two annotators, yielding a Cohen's $\kappa$ of $0.74$, indicating substantial inter-annotator agreement. 
The annotators then met to discuss and resolve all conflicts in accordance with the taxonomy goals of comprehensiveness and applicability across languages and domains.
During this resolution, we exclude instances that are overly vague, purely design-level with no direct performance impact, or exclusively tied to specific applications or languages (\eg{} mobile-specific ``\texttt{Faulty GPS Behavior}'' or Java-specific ``\texttt{Use Static Final for Constants}'').
We retain original context-specific implementations if their underlying inefficiency generalizes across domains. For example, database-specific ``\texttt{Unnecessary Row Retrieval}'' generalizes to fetching unnecessary data in any program. Furthermore, we also retain traditional code smells that inherently introduce resource inefficiencies, as well as low-level hardware and memory behaviors (\eg{} ``Inefficient Array Declaration Order''). This ensures our taxonomy comprehensively spans the full spectrum of inefficiencies.
At the end of this phase, $320$ items and their definitions are retained.

\subsection{Phase 4: Grounded Theory Coding Process}
\label{subsec:coding}
To systematically analyze the $320$ retained items towards taxonomy generation, we adopt a grounded theory coding approach proposed by Strauss and Corbin~\cite{strauss1998basics}, as recommended for qualitative analysis in software engineering research~\cite{braun2006thematic, jahan2025attention, cruzes2011recommended}. 
As shown in Step~\circled{4} of Fig.~\ref{fig:methodology},
 both annotators perform this process independently across two complete rounds, applying three iterative coding steps:
\begin{enumerate}
        \item \textbf{Open coding}: A line-by-line analysis of each extracted item to isolate its fundamental ``root cause'', stripping away application-specific contexts to identify the underlying language-agnostic mechanism. 
        We assign an initial code (\eg{}``\texttt{Fetching never used data}'' or ``\texttt{Using a suboptimal memory-heavy data structure}'') to each instance that describes the technical behavior causing the inefficiency.
        
        \item \textbf{Axial coding}: The initial codes are compared against one another to identify relationships, redundancies, and shared technical contexts. Codes that describe functionally identical root causes under different names are merged, while complex codes that conflate multiple distinct mechanisms are split.
        
        \item \textbf{Selective coding}: 
        The refined codes are abstracted into high-level categories. We examine the conceptual boundaries of the subcategories and group them based on their overarching impact on energy consumption, system resources, and execution flow. For example, codes dealing with unused outputs, repeated identical calculations, and dead branches are grouped under the encompassing category of \textit{Redundant Computation}. 
\end{enumerate}

To ensure clarity and immediate readability, we adopt a bipartite naming convention documented in traditional code smell literature~\cite{suryanarayana2014refactoring,sharma2018survey}.
Each name consists of two components: (1) a \textit{prefix adjective} serving as an inefficiency modifier (\eg{} \textit{Redundant}, \textit{Unnecessary}, \textit{Suboptimal}, \textit{Missing}) that defines how energy is wasted, and (2) a \textit{suffix noun} identifying the specific programmatic construct or behavior involved (\eg{} \textit{Computation}, \textit{Data Structures}, \textit{Control Flow}).

\subsection{Phase 5: Taxonomy Finalization}
In the final phase, \ie{} Step \circled{5} of Fig.~\ref{fig:methodology}, both the annotators collaboratively finalize the taxonomy. This step is conducted jointly and follows a strict top-down resolution order to ensure minimal overlap and maximum comprehensiveness.

\paragraph{Category resolution}
We first reconcile the high-level categories proposed by both annotators to establish the final set of energy smells.

\begin{itemize}
    \item \textbf{Clear one-to-one mapping (category--category):} If both annotators propose categories with similar explanations, we merge them, combine their definitions, and assign a unified name following the naming convention.
    
    \item \textbf{Clear one-to-one mapping (category--sub-category):} If one annotator defines an issue as a category and the other as a sub-category, we jointly evaluate its scope. Issues reflecting a general theme are finalized as categories; issues representing a specific root cause are retained as subcategories.
    
    \item \textbf{No clear mapping:} For categories proposed by only one annotator, we assess granularity. General issues are added as new categories; highly granular issues that qualify as root cause are retained as subcategories. If, after discussion, an issue does not fit either level, it is discarded.
\end{itemize}

After this step, all categories are agreed upon, and the taxonomy contains the final set of energy smells, while root causes remain to be mapped.

\paragraph{Sub-category resolution}
Next, we align the combined list of sub-categories from both annotators.

\begin{itemize}
    \item \textbf{Clear one-to-one mapping:} Subcategories with a clear correspondence are merged. We apply the naming convention, consolidate definitions into a precise explanation, and create an explicit ``example'' column derived from grounded literature. This example reflects the subcategory's behavior and supports validation. Each confirmed subcategory is then assigned to the most appropriate fixed category.
    
    \item \textbf{No clear mapping:} Remaining subcategories are evaluated collaboratively to avoid overlap. Extremely granular codes that apply only in narrow situations are absorbed into an existing subcategory as specific instances. Codes with appropriate granularity and no overlap with existing subcategories are formalized as new subcategories and assigned to the most fitting category.
\end{itemize}

\paragraph{Alignment verification}
After all sub-categories are placed, we perform a final validation pass. We verify that each subcategory, along with its definition, resides under the most appropriate category. If a subcategory shows stronger alignment with a different category based on its definition similarity, it is reassigned.

As a final sanity check, we verify that every retained energy smell is reported in more than one independent source, ensuring that no category relies on a single isolated mention. This finalization process produces a comprehensive, language-agnostic taxonomy consisting of 12 categories (\ie{}, energy smells) and 65 sub-categories (\ie{}, root causes), with no energy smells removed since each was identified in multiple sources.

\subsection{Dataset Creation and Taxonomy Validation}
\label{sec:dataset}

We construct an annotated dataset of energy smells to empirically validate our proposed taxonomy. This process is demonstrated in Step~\circled{6} of Fig.~\ref{fig:methodology}.
Given a pair of functionally equivalent code snippets with distinct energy profiles, an underlying energy smell must account for the difference. Mapping the exact root cause of the inefficiency in these code pairs to our taxonomy confirms the real-world validity of our categories and sub-categories. This mapping also reveals the prevalence of specific energy smells, identifies subcategories that may not appear in practice, and surfaces potential gaps if an inefficiency cannot be mapped.

\subsubsection{Data selection and filtering}
We utilize the Python splits of the Pie-Perf dataset~\cite{madaan2023learning}, which is derived from IBM CodeNet~\cite{ibm_codenet}. Each instance in this dataset contains a problem description, a pair of efficient and inefficient code snippets, and their respective CPU time and memory consumption. We select Pie-Perf because it provides a highly diverse, application-agnostic collection of implementations that effectively captures a wide range of real-world developer habits and algorithmic approaches.
To ensure data integrity, we first verify that both code versions in a pair are functionally correct and yield the same output. Out of $36,857$ initial instances, $21,428$ pairs successfully pass the functional test cases. These pairs form our filtered base dataset.

\subsubsection{Energy profiling and metric extraction}
Pie-Perf inherently focuses only on execution time and memory;
therefore, we must empirically measure the energy consumption of each pair. Following the best practices for accurate energy profiling of software~\cite{rajput2024enhancing}, we execute a rigorous energy measurement pipeline. 
For each of the $21,428$ functionally correct pairs, we randomly select $50$ of the $100$ available test cases. To eliminate hardware state noise, we perform a warm-up run followed by three measured iterations for both the efficient and inefficient snippets. We use the Linux \textit{perf} command-line tool to capture system-wide CPU and RAM energy consumption alongside elapsed execution time. Concurrently, we utilize \textit{time} tool to record the maximum resident set size (RSS), capturing peak physical memory allocation during execution. This process appends six verified metrics (energy, memory, and time for both snippet versions) to each instance.

\subsubsection{Energy-based sample selection}

To focus on pairs with meaningful energy differences, we sort the $21,428$ pairs by the absolute energy difference between their efficient and inefficient versions. We rank the pairs by absolute energy difference and keep the top $3,000$. These pairs account for $98.4$\% of the total energy difference in the dataset, and the cutoff lies beyond the knee of the cumulative-difference curve (rank=$2,123$). In contrast, each excluded pair contributes only 6.8 J on average (versus 84 J at the cutoff). Hence, this threshold provides sufficient coverage to validate all smell categories while ensuring that observed energy differences can be confidently attributed to the identified smells rather than noise or external factors.

\subsubsection{Multi-step LLM classification pipeline}
\label{sec:classification}
Mapping $3,000$ complex algorithmic pairs to specific energy smells requires advanced reasoning, making exhaustive manual classification impractical. Consequently, we delegate the classification task to \texttt{DeepSeek-V3.2 (685B)}~\cite{deepseekai2025deepseekv32}, a frontier model that demonstrates state-of-the-art reasoning capabilities.
We configure the model with a temperature of $0$ for deterministic output, allocate $4,096$ maximum reasoning tokens per step, disable caching, and process each instance independently.

To mitigate priming bias~\cite{chen2024ai} and prevent the model from hallucinating fits to our taxonomy, we divide the classification into a three-step sequential pipeline by designing prompts that operate as follows:
\begin{enumerate}
    \item \textbf{Root cause analysis.} Given the problem description and few-shot examples, the model compares functionally equivalent Python implementations across energy, time, and memory metrics to identify concrete code-level inefficiencies. Crucially, taxonomy labels are \textit{not} provided to ensure unbiased reasoning derived strictly from the code.

    \item \textbf{Category triage.} The model performs a multi-label semantic match, mapping the unbiased root causes extracted in Step 1 to one or more high-level energy smell categories based on provided definitions.

    \item \textbf{Subcategory classification.} Using provided definitions and examples, the model refines Step 2's candidates into precise, multi-label subcategories. To ensure actionable results, only independent, refactoring-relevant root causes are reported: downstream symptoms that disappear when fixing a primary smell are ignored, whereas multiple independent inefficiencies are all captured.
\end{enumerate}

The dataset, along with the reasoning trace output of all the steps, is made available with the dataset~\cite{anonymous_2026_18896365}.

\subsubsection{Dataset evaluation}

To evaluate our classification pipeline, we select a random subset of $100$ samples, and the first two authors independently validate each sample. In Step 1, the model accurately identified the root cause of inefficiency in all $100$ cases without exposure to our taxonomy. Reviewing these explanations revealed no novel inefficiencies outside our defined categories, providing empirical validation for the taxonomy's comprehensiveness. 
Because Step 2 is an intermediate mapping phase, we focused our remaining evaluation on Step 3, which is the final classification step, to ensure the pipeline accurately isolated primary root causes rather than downstream effects. 
The annotators achieved substantial initial agreement (Cohen's $\kappa = 0.65$). 
The annotators then discussed and mutually resolved all conflicting labels to form a final ground-truth dataset. 
Comparing this resolved ground-truth against the pipeline's output yielded an exact-match accuracy of $94$\%. Statistically, a sample size of $100$ drawn from our $3,000$ profiled instances with this observed agreement provides a $95\%$ confidence interval with a margin of error of $\pm 4.6\%$, establishing the mathematical reliability of the classification. The $6$ misclassifications primarily involved highly complex, multifaceted algorithmic flaws where isolating a single root cause is inherently ambiguous, confirming that the pipeline reliably isolates core energy smells without heavily over-predicting downstream symptoms. Also, scoring the predicted (instance, subcategory) pairs against the resolved labels yields micro precision/recall/F1 of $0.97$, and exact-match rises to $96\%$ at the category level, demonstrating the LLM's highly-accurate decision-making. 

\definecolor{catC1}{HTML}{FCE4D6}
\definecolor{catC2}{HTML}{DDEBF7}
\definecolor{catC3}{HTML}{E2EFDA}
\definecolor{catC4}{HTML}{F9E79F}
\definecolor{catC5}{HTML}{FFF2CC}
\definecolor{catC6}{HTML}{E2D9F3}
\definecolor{catC7}{HTML}{D9E2F3}
\definecolor{catC8}{HTML}{D5F5E3}
\definecolor{catC9}{HTML}{D6DCE4}
\definecolor{catC10}{HTML}{FADBD8}
\definecolor{catC11}{HTML}{FDEBD0}
\definecolor{catC12}{HTML}{D4E6F1}

\begin{table*}[tbp]
\centering
\vspace{-2mm}
\caption{Energy smell taxonomy: 12 energy smells and 65 root causes.}
\vspace{-1mm}
\label{tab:taxonomy-overview}
\footnotesize
\renewcommand{\arraystretch}{1.1}
\setlength{\tabcolsep}{5pt}
\begin{tabular}
{@{}p{0.14\textwidth}p{0.27\textwidth}p{0.55\textwidth}@{}}
\toprule
\textbf{Category} & \textbf{Root Cause (ID)} & \textbf{Description} \\
\midrule

% CATEGORY 1
\rowcolor{catC1} \textbf{C1: Redundant} & Dead Code (C1.S1) & Code with \textbf{unused results} or \textbf{unreachable paths}. \\
\rowcolor{catC1} \textbf{Computation} & Redundant Assignment (C1.S2) & Statements leaving \textbf{state unchanged} (self-assignment, overwriting unread values). \\
\rowcolor{catC1} \textit{30 occurrence} & Redundant Control Flow (C1.S3) & Control flow that \textbf{does not alter execution} regardless of branch taken. \\
\rowcolor{catC1} \textit{13 sources ($\bar{c} \approx 123$)} & Repeated Computation (C1.S4) & \textbf{Identical calculations repeated} in the same scope without data changes. \\
\rowcolor{catC1} & Resultless Computation (C1.S5) & Operations producing \textbf{no subsequently used output}. \\
\rowcolor{catC1} & Unnecessary Initialization (C1.S6) & Setup on paths where results are \textbf{unused or overwritten}. \\
\rowcolor{catC1} & Unnecessary Variable (C1.S7) & Intermediate variables whose \textbf{overhead exceeds reuse} utility. \\
\rowcolor{catC1} & Excessive Runtime-Mgmt (C1.S8) & \textbf{Excessive housekeeping} or logging invocations wasting energy. \\

% CATEGORY 2
\rowcolor{catC2} \textbf{C2: Unnecessary} & Unnecessary Delegation (C2.S1) & \textbf{Wrappers or accessors} adding overhead without benefit. \\
\rowcolor{catC2} \textbf{Call Overhead} & Missing Static Declaration (C2.S2) & \textbf{Instance methods} for operations independent of instance state. \\
\rowcolor{catC2} \textit{23 occurrence \newline 15 sources ($\bar{c} \approx 62$)} & Excessive Modularization (C2.S3) & Hot paths \textbf{over-split} into tiny layers where call overhead rivals useful work. \\

% CATEGORY 3
\rowcolor{catC3} \textbf{C3: Inefficient} & Inefficient Iteration Construct (C3.S1) & Iteration forms with \textbf{avoidable overhead} when efficient alternatives exist. \\
\rowcolor{catC3} \textbf{Iteration Patterns} & Recomputing Loop-Invariant (C3.S2) & \textbf{Loop-invariant values} recomputed inside loops instead of once outside. \\
\rowcolor{catC3} \textit{42 occurrence} & Inefficient Per-Iteration Setup (C3.S3) & \textbf{Heavy initialization} inside loops that could be pre-computed. \\
\rowcolor{catC3} \textit{22 sources ($\bar{c} \approx 66$)} & Inefficient Nested Iteration (C3.S4) & \textbf{Nested loops} where an equivalent $O(n)$ approach exists. \\
\rowcolor{catC3} & Missing Loop Early Exit (C3.S5) & Failing to \textbf{exit early} after determining the required result. \\
\rowcolor{catC3} & Unfiltered Bulk Iteration (C3.S6) & Processing \textbf{entire collections} when only a subset is needed. \\
\rowcolor{catC3} & Inefficient Array Mutation (C3.S7) & \textbf{Modifying collections during iteration}, causing re-indexing or anomalies. \\

% CATEGORY 4
\rowcolor{catC4} \textbf{C4: Inefficient} & Poor Short-Circuit Ordering (C4.S1) & \textbf{Poor operand ordering} preventing optimal short-circuit evaluation. \\
\rowcolor{catC4} \textbf{Control Flow} & Redundant Conditional (C4.S2) & Evaluating \textbf{statically known} or \textbf{unchanged} conditions. \\
\rowcolor{catC4} \textit{22 occurrence} & Inefficient Conditional Nesting (C4.S3) & \textbf{Deep nesting} hindering early returns and branch optimization. \\
\rowcolor{catC4} \textit{14 sources ($\bar{c} \approx 45$)} & Missing Else-If (C4.S4) & \textbf{Independent conditionals} for mutually exclusive branches. \\
\rowcolor{catC4} & Expensive Comparison (C4.S5) & Costly \textbf{reflection or type introspection} instead of simpler comparisons. \\
\rowcolor{catC4} & Expensive Exception Flow (C4.S6) & \textbf{Exceptions for expected control flow}, incurring high object creation costs. \\
\rowcolor{catC4} & Missing Edge-Case Guards (C4.S7) & Missing \textbf{early guards} for trivial inputs, forcing expensive general-case processing. \\
\rowcolor{catC4} & Non-Idiomatic Condition (C4.S8) & \textbf{Non-idiomatic} conditional patterns missing runtime optimizations. \\

% CATEGORY 5
\rowcolor{catC5} \textbf{C5: Suboptimal} & Inefficient Structure Choice (C5.S1) & Data structures with \textbf{suboptimal complexity} for the dominant operation. \\
\rowcolor{catC5} \textbf{Data Structures} & Missing Helper Type (C5.S2) & Missing \textbf{auxiliary structures} (caches, indices) to optimize repeated lookups. \\
\rowcolor{catC5} \textit{34 occurrence} & Over-Provisioned Data Type (C5.S3) & \textbf{Heavier types} than data constraints necessitate. \\
\rowcolor{catC5} \textit{21 sources ($\bar{c} \approx 95$)} & Unnecessary Representation (C5.S4) & \textbf{Format conversions} without net efficiency gains. \\

% CATEGORY 6
\rowcolor{catC6} \textbf{C6: Unnecessary} & Unnecessary Object (C6.S1) & \textbf{Excessive, short-lived, or duplicate} objects increasing GC pressure. \\
\rowcolor{catC6} \textbf{Memory Usage} & Unnecessary Copying (C6.S2) & \textbf{Copying} when views, references, or in-place operations suffice. \\
\rowcolor{catC6} \textit{31 occurrence} & Unnecessary Materialization (C6.S3) & Fully \textbf{materializing} intermediate data instead of lazy evaluation. \\
\rowcolor{catC6} \textit{20 sources ($\bar{c} \approx 74$)} & Oversized Data Retaining (C6.S4) & Retaining \textbf{full structures} when smaller representations (IDs) suffice. \\
\rowcolor{catC6} & Over-Allocation (C6.S5) & \textbf{Over-allocating} buffers, wasting memory and increasing cache pressure. \\
\rowcolor{catC6} & Leaked Resource Handles (C6.S6) & Failing to \textbf{release external resources} (files, sockets, connections). \\
\rowcolor{catC6} & Leaking Mutable Defaults (C6.S7) & \textbf{Mutable default parameters} causing silent persistence and growth. \\

% CATEGORY 7
\rowcolor{catC7} \textbf{C7: Suboptimal} & Suboptimal Algorithm Choice (C7.S1) & Algorithms with \textbf{worse complexity} when efficient alternatives exist. \\
\rowcolor{catC7} \textbf{Algorithmic} & Inefficient Decomposition (C7.S2) & \textbf{Redundant or poorly-ordered} subcomputations. \\
\rowcolor{catC7} \textit{17 occurrence} & Avoidable Recursion (C7.S3) & \textbf{Recursion} where iterative solution is more efficient. \\
\rowcolor{catC7} \textit{12 sources ($\bar{c} \approx 73$)} & Inefficient Operation Ordering (C7.S4) & \textbf{Expensive operations first} when cheaper ones could prune input. \\
\rowcolor{catC7} & Unsimplified Operation (C7.S5) & Heavy operations on inputs that could be \textbf{pre-reduced or simplified}. \\

% CATEGORY 8
\rowcolor{catC8} \textbf{C8: Missing} & Missing Memoization Cache (C8.S1) & \textbf{Recomputing} deterministic results instead of memoizing. \\
\rowcolor{catC8} \textbf{Reuse} & Missing Derived-Value Reuse (C8.S2) & \textbf{Recreating} expensive artifacts (regex, SQL) on every use. \\
\rowcolor{catC8} \textit{21 occurrence} & Missing Local Lookup Caching (C8.S3) & Repeating \textbf{global/attribute lookups} in hot code instead of caching locally. \\
\rowcolor{catC8} \textit{15 sources ($\bar{c} \approx 51$)} & Redundant Data Fetching (C8.S4) & \textbf{Repeatedly fetching} unchanged external data instead of caching. \\

% CATEGORY 9
\rowcolor{catC9} \textbf{C9: Inefficient} & Fragmented I/O Calls (C9.S1) & \textbf{Unbatched} small I/O or network operations. \\
\rowcolor{catC9} \textbf{External Access} & Oversized Data Retrieval (C9.S2) & Fetching \textbf{more data than consumed} by application logic. \\
\rowcolor{catC9} \textit{39 occurrence} & Inefficient Retrieval Paths (C9.S3) & \textbf{Multiple retrieval steps} instead of a single joined query. \\
\rowcolor{catC9} \textit{19 sources ($\bar{c} \approx 114$)} & Expensive Query Pattern (C9.S4) & Query structures \textbf{inherently costly} for database engines. \\

% CATEGORY 10
\rowcolor{catC10} \textbf{C10: Underused} & Skipping Optimized Built-ins (C10.S1) & \textbf{Manual reimplementation} of optimized built-in functionality. \\
\rowcolor{catC10} \textbf{Language Primitives} & Inefficient Built-in Choice (C10.S2) & Using constructs when \textbf{faster equivalents} exist. \\
\rowcolor{catC10} \textit{45 occurrence} & Missing Bulk Primitive Usage (C10.S3) & \textbf{Per-element operations} instead of vectorized/batched primitives. \\
\rowcolor{catC10} \textit{16 sources ($\bar{c} \approx 49$)} & Inefficient String Concat (C10.S4) & \textbf{Repeated concatenation} of immutable strings, causing excessive allocation. \\
\rowcolor{catC10} & Inefficient Scope Lookup (C10.S5) & Variables in scopes with \textbf{unnecessary lookup overhead}. \\
\rowcolor{catC10} & Expensive Operator Overloads (C10.S6) & \textbf{Expensive overloaded methods} invoked via operator syntax on non-primitives. \\

% CATEGORY 11
\rowcolor{catC11} \textbf{C11: Inefficient} & Excessive Lock Contention (C11.S1) & \textbf{Overly broad synchronization} serializing parallelizable work. \\
\rowcolor{catC11} \textbf{Concurrency} & Forced Serial Bottleneck (C11.S2) & \textbf{Single-file execution} through shared resource, negating parallelism. \\
\rowcolor{catC11} \textit{9 occurrence} & Leaked Background Threads (C11.S3) & \textbf{Leaked threads/tasks} without clean termination. \\
\rowcolor{catC11} \textit{9 sources ($\bar{c} \approx 142$)} & Blocking The Main Thread (C11.S4) & \textbf{Blocking main thread} with long-running CPU or I/O work. \\
\rowcolor{catC11} & Missed Parallelism (C11.S5) & \textbf{Serial execution} of independent work, underutilizing resources. \\

% CATEGORY 12
\rowcolor{catC12} \textbf{C12: Poor Hardware} & Inefficient Large-Stride (C12.S1) & \textbf{Large stride} array traversal causing frequent cache misses. \\
\rowcolor{catC12} \textbf{Locality} & Sparse Element Access (C12.S2) & \textbf{Scattered access} in large structures, preventing efficient cache use. \\
\rowcolor{catC12} \textit{5 occurrence} & Unpredictable Branches (C12.S3) & \textbf{Data-dependent branching} in tight loops, defeating branch prediction. \\
\rowcolor{catC12} \textit{2 sources ($\bar{c} \approx 156$)} & Inefficient Array Declaration (C12.S4) & \textbf{Hot buffers after cold ones}, resulting in poor cache alignment. \\
\bottomrule
\end{tabular}
\vspace{-2mm}
\end{table*}
\section{Energy Smell Taxonomy}
\label{sec:taxonomy}

Table~\ref{tab:taxonomy-overview} presents the full taxonomy of twelve energy smells and $65$ root causes, where each energy smell tells a developer \textit{what kind} of inefficiency is present and each root cause tells them \textit{what to detect and how to fix it}. We empirically validate the taxonomy by classifying $3,000$ code pairs from the \textit{Pie-Perf} dataset using the three-step LLM pipeline (Sections~\ref{sec:dataset},~\ref{sec:classification}). 
In this section, we introduce the twelve categories organized into four thematic groups, explain category boundary decisions, demonstrate classification on real code, and report empirical coverage.

\paragraph{Wasted work (C1, C2, C4)}
\textit{Redundant Computation}~(C1) covers code whose results are never consumed or that repeats computation already done: dead code, redundant assignments, repeated expressions, and unnecessary initializations.
\textit{Unnecessary Call Overhead}~(C2) targets function calls, method delegation, and dynamic dispatch that add cost without proportional benefits.
\textit{Inefficient Control Flow}~(C4) addresses branching logic that performs unnecessary checks, uses suboptimal evaluation order, or prevents runtime optimization.

\paragraph{Iteration and loop structure (C3)} \textit{Inefficient Iteration Patterns}~(C3) captures loops that do more per-iteration work than necessary, iterate more times than required, or fail to terminate early. Due to their repetitive nature, issues in this category often amplifies issues from other categories
(Section~\ref{subsec:rationale}).

\paragraph{Data and memory (C5, C6)}
\textit{Suboptimal Data Structures}~(C5) covers choosing containers that do not match the workload's access pattern, such as using a list for frequent membership queries when a set provides constant-time lookup.
\textit{Unnecessary Memory Usage}~(C6) captures allocating, copying, or retaining memory beyond what the workload needs, including unnecessary object creation, premature materialization of lazy sequences, and over-allocation.

\paragraph{Algorithm and reuse (C7, C8)} \textit{Suboptimal Algorithmic}~(C7) covers choosing an algorithm or decomposition strategy with higher complexity than the problem requires.
\textit{Missing Reuse}~(C8) addresses failing to store and reuse results of expensive computations across calls, including missing memoization and redundant data fetching.

\paragraph{External systems and concurrency (C9, C11)} \textit{Inefficient External Data Access}~(C9) targets inefficient interaction with databases, file systems, and APIs.
\textit{Inefficient Concurrency Management}~(C11) captures thread and task misuse: excessive lock contention, serial bottlenecks, and missed parallelism opportunities.

\paragraph{Language and hardware (C10, C12)} \textit{Underused Language Primitives}~(C10) covers not using available optimized built-in functions or runtime constructs.
\textit{Poor Hardware Locality}~(C12) targets cache misses, branch misprediction, and suboptimal memory access patterns relevant when code interacts with contiguous-memory libraries or native extensions.

\begin{figure}[t]
    \centering
    \includegraphics[width=1\linewidth]{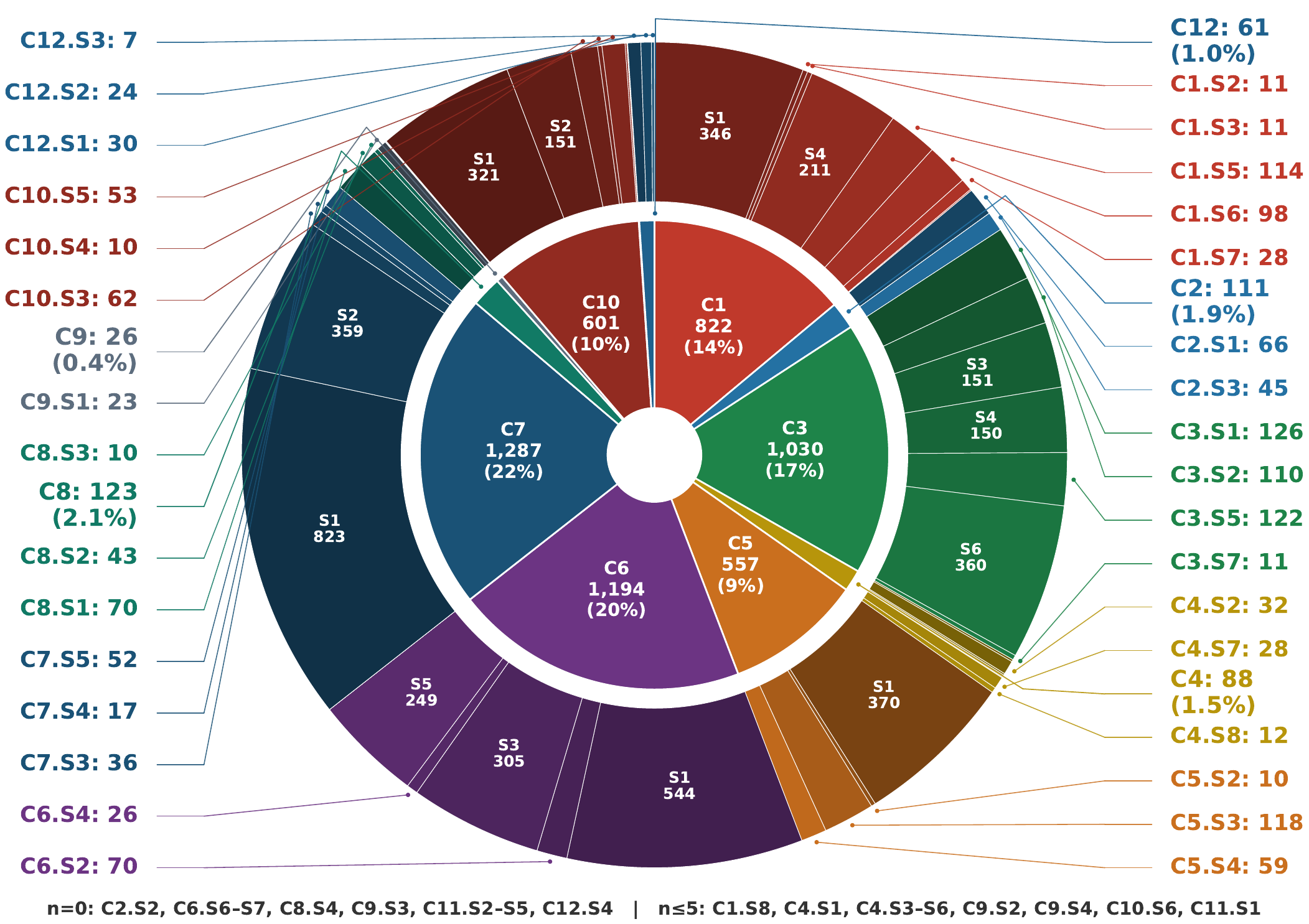}
    \caption{Energy smells and root causes distribution}
    
    \label{fig:dist}
\end{figure}

\subsection{Empirical Coverage}
\label{subsec:coverage}
As detailed in Section~\ref{sec:classification}, applying our classification pipeline to the Pie-Perf dataset yielded 3,000 successfully categorized code pairs.
We observe $55$ ($84.6$\%) out of the $65$ subcategories at least once in the $3,000$ classified samples. 
Fig.~\ref{fig:dist} depicts
the prevalence of all subcategories. The frequency distribution reflects strong algorithmic stress-testing in the dataset, dominated by \textit{Suboptimal Algorithm Choice} (C7.S1, $822$) and \textit{Inefficient Problem Decomposition} (C7.S2, $359$). The high occurrence of \textit{Unnecessary Object Creation} (C6.S1, $543$) and \textit{Inefficient Data Structure Choice} (C5.S1, $370$) suggests developers often prioritize quick logical correctness over memory-efficient abstractions. Similarly, frequent \textit{Skipping of Optimized Built-ins} (C10.S1, $321$) indicates a tendency to reimplement logic instead of using native library functions, reflecting limited language-specific performance awareness. Finally, the prevalence of \textit{Unfiltered Bulk Iteration} (C3.S6, $360$) and \textit{Unnecessary Data Materialization} (C6.S3, $304$) reveals a bias toward eager execution rather than energy-efficient lazy evaluation or generator pipelines. 

The absent subcategories can be grouped in three clusters: concurrency patterns ($4$ out of $5$ C11 subcategories), requiring multi-threaded execution absent from single-threaded competitive code; external data access (C8.S4, C9.S3), requiring database or I/O interaction; and resource-lifecycle patterns (C6.S6, C6.S7), requiring persistent connections or multi-call objects. C2.S2 and C12.S4 are also absent due to Python's dispatch model and C-extension specificity, respectively. Every absent subcategory maps to a structural feature absent from the dataset's domain, not to a gap in the taxonomy. 

\subsection{Literature Grounding of the Energy Smells}

\label{subsec:grounding}
Our taxonomy is intended to be comprehensive by construction. Rather than targeting a single domain, it aggregates the inefficiencies reported across all sources admitted by our search and selection criteria, including the existing energy and performance anti-pattern catalogs, which we coded as input sources so that their categories are absorbed into our taxonomy. To avoid missing relevant smells, we cast a deliberately wide net: a broad keyword query, the inclusion of performance-related inefficiencies alongside energy-specific ones, and exhaustive forward and backward snowballing over the retained studies. This breadth directly addresses the limitations of prior work discussed in Sections~\ref{sec:intro} and~\ref{sec:relatedwork}, where existing catalogs are often confined to a single domain or omit several classes of energy smells, and it gives us reasonable confidence that the taxonomy is wide enough to capture the important inefficiencies.

Beyond this breadth of coverage, our taxonomy is also grounded by construction: most of the patterns are drawn from peer-reviewed studies, so each smell inherits the validation already conferred by its originating venue. Whereas the coverage analysis in Section~\ref{subsec:coverage} examines whether the smells manifest in real code, we now consider a complementary question, namely the extent to which each category is independently attested across the surveyed literature. For every energy smell, Table~\ref{tab:taxonomy-overview} reports the number of the 320 retained patterns it subsumes, the number of \emph{distinct} sources that describe it, and the average citation count of those sources ($\bar{c}$). The recurrence of a given inefficiency across multiple independent sources provides corroborating evidence that it constitutes a generalizable concern rather than an isolated observation. Furthermore, $\bar{c}$ serves as a proxy for the scholarly influence and visibility of the supporting sources, providing an additional indication of the sources credibility. We confine this analysis to the category level: the root causes are axial refinements of the same corroborated sources, and their presence in practice is already established by the coverage results $\approx 85\%$. 

For C1--C10, the supporting evidence is substantial. Each category is reported by at least 12 distinct sources, and the corresponding sources are well cited on average, with mean counts ranging from approximately 45 to 123. This breadth is expected, as these categories capture general-purpose inefficiencies, including redundant computation, poorly chosen data structures and algorithms, unnecessary memory traffic, and the under-use of optimized language primitives, that recur across application domains and, accordingly, throughout the literature.
The concurrency (C11) and hardware-locality (C12) smells are supported by fewer sources. C11 is described by nine independent sources, each contributing a single pattern. Concurrency inefficiencies receive comparatively little attention in the literature, as parallelism is frequently presumed beneficial despite its potential to increase power consumption, and such code probably arises less often in routine development. Although this category is supported by fewer sources than others, the quality and influence of those sources are substantial ($\bar{c} \approx 142$), justifying its inclusion in the taxonomy. Moreover, C12 is the least represented, resting on two sources, both of which are nonetheless highly cited (54 and 258). Similar to concurrency, hardware-locality concerns constitute a more specialized subject addressed by a narrower body of work; we retain C12 because energy is ultimately dissipated at the hardware level, and the standing of these peer-reviewed sources affords reasonable confidence in this energy smell.

\subsection{Taxonomy Characterization}
\label{subsec:characterization}
The root causes are not uniformly distributed across categories. At the structural level, C1, C4 (eight root causes each), and C3 ($7$) are the largest, reflecting the density of distinct mechanisms at the implementation level, while C2 ($3$) and C5 ($4$) are smaller because they address fewer, more structurally distinct mechanisms. The empirical prevalence from the $3,000$ classified samples, however, reveals a different ranking. C7 (only five root causes) is the most prevalent category at $41.6$\% of samples, followed by C6 at $38.9$\% and C3 at $33.3$\%. At the other end, C11 and C9 each appear in fewer than $1$\% of samples, which is expected given the dataset's single-threaded, in-memory domain. 

The taxonomy spans three abstraction levels. \textit{Implementation-level} categories (C1, C2, C3, C4, C6, C10) target small code snippets fixable by editing a few lines. \textit{Design-level} categories (C5, C7, C8) require choosing a different data structure, algorithm, or caching strategy. \textit{Architecture-level} categories (C9, C11) involve interaction with external systems or concurrency models; C12 is syntactically implementation-level but operates at the hardware level. The empirical data confirms this distinction: design-level smells produce higher median energy savings per fix ($1,537$~J, $N$=$1,725$) than implementation-level smells ($1,146$~J, $N$=$2,476$), consistent with the expectation that deeper design fixes yield larger gains. Architecture-level ($N$=$28$) and hardware-level ($N$=$61$) smells show lower medians ($415$~J and $561$~J), though this partly reflects their under-representation in competitive programming code.

\subsection{Category Design and Boundary Decisions}
\label{subsec:rationale}

Designing the taxonomy required resolving cases where the same inefficiency could plausibly belong to more than one category. The guiding principles are to minimize overlap so that each inefficiency maps to exactly one primary root cause, and to ensure that the category and subcategory 
boundaries support distinct refactoring actions. Below we explain the non-obvious boundary decisions. Where applicable, we supplement these with disambiguation evidence from the classification pipeline, in which a sample initially matched multiple candidate categories but was resolved to one based on the taxonomy rules.

\paragraph{Redundant vs. Repeated computation  (C1~vs.~C8)} Both categories involve redundant work, but they differ in scope. C1 covers computations whose result has no impact on the final functional outcome: dead code, self-assignments, and unused initializations. The fix is deletion, since the work never needed to happen. C8 covers computation that is necessary and produces a useful result, but the program recomputes it from scratch on every call instead of caching it. The fix is not deletion but adding a reuse mechanism, such as memoization or precomputation. This distinction carries over to tooling: detecting C1 requires only intra-procedural analysis to identify unused results, while detecting C8 requires tracking repeated calls with identical inputs across invocations.

\paragraph{Loops as amplifiers (C3~vs.~other categories)} Loops multiply any inefficiency by the iteration count, which is why several C3 root causes have counterparts elsewhere. \textit{Per-Iteration Setup}~(C3.S3) captures waste from creating objects \textit{inside a loop body} that should be hoisted out; \textit{Missing Derived-Value Reuse}~(C8.S2) captures the same artifact being recreated \textit{across function calls}. Similarly, a hand-written summation loop could be classified as C3.S1 or C10.S1; we assign C10.S1 because the root cause is not using the available built-in function.

\paragraph{Data structures vs. algorithms (C5~vs.~C7)} C5 addresses choosing the wrong \textit{container} for the workload's access pattern; C7 addresses choosing the wrong \textit{algorithm} or decomposition strategy. The two require different refactoring actions even when they co-occur.

\subsection{Overlap and Classification Rules}
\label{subsec:overlap}

A single code fragment can trigger multiple energy smells simultaneously. In the classified dataset, $71$\% of the $3,000$ samples received two or more labels ($48$\% received exactly two, $20.4$\% received three), confirming that multi-label classification is the norm, not an edge case. Samples with multiple labels exhibit significantly higher energy waste: multi-label cases have a median energy savings of $1,659$~J compared to $549$~J for single-label cases (Mann-Whitney $U$, $p < 0.0001$), a $3.0$$\times$ difference. This indicates that energy waste compounds when multiple inefficiencies interact.

To handle this overlap during classification, we established a strict decision rule: \textit{assign the root cause whose fix directly eliminates the primary inefficiency, rather than tagging downstream consequences or amplifiers.} Out of the $3,000$ samples, $765$ ($25.5$\%) initially had multiple candidate categories in Step~2 but were resolved to a single, deeper root cause using this rule.

\subsection{Classification Walkthrough}
\label{subsec:walkthrough}
We present two examples from the dataset
to illustrate how the taxonomy classifies real code.

\paragraph{Example 1: Unused import as dead code (Dataset \#$4327$)} The inefficient version contains \texttt{import numpy as np} but never references NumPy. Importing this library is costly because it loads large C-extensions and other initialization, which consumes CPU and memory before any application logic runs.
Removing this single line saves $4,036$~J ($13.5$$\times$ ratio). The LLM considered both C1 (dead code) and C6 (memory overhead from loading NumPy's C extensions). The final classification is C1.S1 (Dead Code): the root cause is that the import should not exist; the memory overhead is a consequence, not an independent problem. 

\paragraph{Example 2: Loop mutation amplifying a suboptimal algorithm (Dataset \#$14434$)} 
A Sieve of Eratosthenes implementation calls \texttt{list.pop()} inside a while loop, producing $O(n^2)$ complexity from repeated element shifting. The efficient version uses iterative primality testing, saving $23,446$~J ($67$$\times$ ratio). This sample receives two labels: C3.S7 (array mutation) and C7.S1 (suboptimal algorithm). Neither identification is complete independently: the algorithm choice is the root cause, but the in-loop mutation is the mechanism that makes it quadratic. Notably, C3.S7 appears only $11$ times but has the highest mean savings ($9,000$~J), showing that rare smells can be costly when present.

\section{Discussion}
\label{sec:discussion}

This section examines the empirical findings from the $3,000$ classified code pairs and discusses their implications for practitioners and researchers.

\paragraph{Frequency and impact do not align}
The distribution of energy savings in our dataset is highly right-skewed, exhibiting a mean of $2,561$~J compared to a much lower median of $1,081$~J (a $2.4$$\times$ ratio). With a skewness of 3.4 and an extreme range ($84$~J to $38,416$~J), the data demonstrates a clear Pareto-like effect: a small fraction of severe inefficiencies accounts for a disproportionate majority of the total energy waste. 
Crucially, the frequency of an energy smell does not dictate its impact. For example, C7 is the most common category ($41.6$\%) but its median savings ($1,119$~J) rank only third. In contrast, C5 ($18.0$\%) and C6 ($38.9$\%) yield the highest median savings at $3,989$~J and $3,670$~J, respectively. This impact is magnified in the extreme upper tail: when isolating the top $25$\% most wasteful code pairs (Q4, savings above $3,996$~J), C6 and C5 dominate, appearing in $57.9$\% and $29.2$\% of samples ($\chi^2$, $p < 0.001$). Minor issues like C4 and C2 are nearly absent from this high-impact tier. 
This mismatch arises because C5 and C6 heavily affect the memory subsystem. Large unnecessary allocations increase DRAM traffic, elevate garbage collection pressure, and raise power draw. While an algorithmic smell primarily extends execution time, a memory smell extends both time and power, producing exponentially higher savings per fix. The practical implication for software tooling is clear: \textbf{energy smell detectors must prioritize warnings by category impact rather than mere frequency.} A single C5 or C6 finding is worth substantially more energy than ten C4 findings (median $\approx307$~J).

\paragraph{Co-occurrence reveals C6 as a structural hub} The five most common co-occurring pairs are C3+C7 ($426$ samples), C1+C6 ($372$), C6+C7 ($341$), C5+C6 ($303$), and C3+C6 ($295$). C6 appears in four of the top five and co-occurs significantly with $8$ of $11$ other categories (Fisher's exact test, $p < 0.05$). Bad algorithms over-allocate proportionally to wasted work; redundant computations create persisting temporary objects; wrong data structures are themselves over-allocation; inefficient iteration forces intermediate collections.

\paragraph{Energy is not a proxy for time}
Three layers of evidence support an energy-specific taxonomy beyond performance taxonomies. First, power draw is not constant: it ranges from $60.5$~W to $147.3$~W (Coefficient of Variation (CV) of $24.2$\%) for the inefficient code. Memory-intensive categories draw significantly more power: C5 has a mean power ratio of $1.267$ and C6 of $1.186$, compared to $1.056$ for C7 and $1.035$ for C8. Fixing a data structure smell reduces both time and power, a double benefit invisible to time-only analysis. Second, $15$ samples show improved time with increased energy, all involving NumPy vectorization: faster wall time but SIMD activation at higher power (\eg{} $78$~W to $135$~W), producing up to $64$\% energy increase. Conversely, $5$ samples show improved energy with worse time, driven by power reduction alone. These $20$ cases confirm that time and energy improvements are distinct axes. Third, in $69.1$\% of samples, the energy ratio ($E_{v0}/E_{v1}$) exceeds the time ratio ($T_{v0}/T_{v1}$), meaning naive estimation using $E = \text{constant} \times T$ systematically underestimates energy waste because inefficient code tends to draw more power in addition to running longer.

\subsection{Implications}
\label{sec:implications}

As Green Software Engineering matures, a critical gap remains between measuring energy waste and actively refactoring code to eliminate it. Our two-level taxonomy bridges this gap by formalizing the path from measurement to remediation. By establishing a shared vocabulary and an empirical foundation, this work streamlines the efforts towards the development of automated tools and integrates energy awareness directly into standard workflows.

Our taxonomy equips \textbf{software developers} with a universal language for energy efficiency, where the high-level \textit{energy smell} serves as a diagnostic signal, while the \textit{root cause} provides an immediately actionable refactoring path that optimize resource usage, execution speed, and client battery life. For \textbf{industry}, organizations can integrate energy smell detection directly into CI/CD pipelines as a proactive safeguard against wasteful code, allowing companies to quantify software quality, benchmark products against industry standards, and significantly reduce server electricity costs while positioning themselves as environmentally responsible. Furthermore, \textbf{AI model developers} can utilize our profiled dataset as a foundation for downstream tools, enabling the supervised training of lightweight, real-time IDE linters and using verified energy, time, and memory metrics, along with the reasoning traces for preference-based alignment (e.g., DPO~\cite{rafailov2023direct}) to teach generative LLMs to synthesize energy-efficient code. Finally, this work opens numerous avenues for \textbf{researchers} to conduct large-scale empirical studies, tracking longitudinal software evolution to investigate prevalence of energy smells, and develop novel metric to quantify them.

\section{Related Work}
\label{sec:relatedwork}

\textbf{Software energy concern in general:}
Software energy consumption has long been a research focus.
Early work by Mehta \etal{}~\cite{mehta1997techniques} explores compiler-level optimizations—such as improved register allocation, loop unrolling, software pipelining, and recursion elimination—to enhance energy efficiency. Similarly, Sinha \etal{}~\cite{sinha2000energy} develop software energy models, showing that restructuring algorithms can significantly improve energy-quality scalability in resource-constrained systems.
Recent studies highlight software’s role in managing hardware power states~\cite{ournani2020reducing,georgiou2017iot}. In fact, Georgiou \etal{}~\cite{georgiou2017iot} advocate incorporation of energy efficient practices in the embedded software systems.
Noureddine \etal{}~\cite{noureddine2015monitoring} introduce a fine-grained runtime framework for estimating Java energy use to detect software energy hotspots, and Georgiou \etal{}~\cite{georgiou2019software} survey techniques, emphasizing best practices and continuous monitoring for improved energy efficiency of software.

\textbf{Impact of software design on energy}:
A distinct line of studies analyzes the effect of software design—such as design patterns~\cite{maleki2017understanding}, code smells~\cite{palomba2019impact}, and refactoring techniques~\cite{pinto2015refactoring}—on software energy consumption.
Recent systematic literature reviews~\cite{poy2025impact,connolly2025software} comprehensively explore this domain, identifying exactly which code smells, refactoring methods, and design patterns positively or negatively impact energy consumption.

\textbf{Energy smells:}
Although software design influences energy use, it mainly targets maintainability. 
We further divide the studies focusing on energy smells in two broad categories.

\subsubsection{Energy anti-pattern definition}
Brandolese \etal{}~\cite{brandolese2002impact} study source-level C transformations for reducing power, identifying anti-patterns such as \textit{oversized loop bodies} and \textit{inefficient array access}. Similarly, Hopfner \etal{}~\cite{hopfner2010towards} introduce a resource substitution matrix for networking, identifying anti-patterns such as \textit{transmitting uncompressed data} or \textit{neglecting local caching}. Many studies in this domain focus specifically on Android applications, as energy directly affects both the user experience and the environment~\cite{carette2017investigating,palomba2019impact}. Gottschalk \etal{}~\cite{gottschalk2012removing} identify the first set of Android energy code smells through a literature review, cataloging patterns such as \textit{binding resources too early, loop bug, immortality bug} alongside their detection and refactoring methods. Similarly, Li \etal{}~\cite{li2014investigation} demonstrate that bundling network packets and optimizing array length reads effectively lower mobile energy consumption. Additionally,
energy efficiency is also critical for battery-powered embedded systems with limited resources~\cite{guo2021survey}. Vetro \etal{}~\cite{vetro2013definition} present the first validated catalog of energy smells for embedded systems, empirically comparing smelly and refactored code. Their catalog includes anti-patterns such as \textit{parameter by value and dead local store}.

In broader domains, GreenForLoops~\cite{gurung2024static} examines Java energy efficiency, identifying loop-related smells such as \textit{array initialization}, \textit{string concatenation}, and \textit{unnecessary nested loops}. Pereira \etal{}~\cite{pereira2016influence} and Oliveira \etal{}~\cite{oliveira2019recommending} analyze the energy impact of alternative Java data structures, showing benefits of optimal substitutions. In the web domain, Rani \etal{}~\cite{rani2024energy} port Android energy anti-patterns to web applications, demonstrating that \textit{reduce size}, \textit{cache}, and \textit{batch operations} improve efficiency.

\subsubsection{Energy smells detection}
Several tools target energy inefficiencies, predominantly in Android. EcoAndroid~\cite{ribeiro2021ecoandroid} automates energy-efficient Java refactorings based on existing literature. Prestat \etal{}~\cite{prestat2022empirical} evaluate static-analysis tools such as PAPRIKA~\cite{hecht2015detecting} and ADOCTOR~\cite{palomba2017lightweight} for detecting mobile resource issues, while ecoCode~\cite{le2022ecocode} broadens prior work by defining novel Android energy smells and providing a dedicated detection tool.

\textbf{Performance anti-patterns:}
Given the relationship between performance and energy,
examining performance-specific inefficiencies is crucial. 
Lyu \etal{}~\cite{lyu2019quantifying} define SQL performance anti-patterns, demonstrating how these database-level issues degrade both runtime and energy on mobile devices. Exploring general software, Zhao \etal{}~\cite{zhao2020performance} analyze 192 real-life performance issues in Java projects, utilizing grounded theory to categorize bottlenecks into eight root causes (\eg{} \textit{inefficient data structures}) and map them to optimization solutions. At the statement level, Tao \etal{}~\cite{tao2024beyond} profile 250 student assignments to extract 27 recurring inefficiencies.
Similarly, Dargan \etal{}~\cite{dargan2025my} analyze slow Python submissions, deriving a taxonomy that groups over 750 efficiency bugs into three primary root causes (\textit{superfluous computation, suboptimal data structures}, and \textit{suboptimal algorithms}) spanning twelve specific subcategories.

\section{Threats to Validity}
\label{sec:threats}

\textbf{Construct validity.} The LLM classification pipeline may introduce misclassification or bias. We mitigate this by keeping the first step independent of the taxonomy to capture genuine root causes, validating on 100 random instances (94\% accuracy), and relying on a large sample size (3,000) for statistical robustness. Energy measurement on a single hardware configuration is another construct threat; we follow established best practices~\cite{rajput2024enhancing} to minimize noise, though absolute joule values may vary across architectures.

\textbf{Internal validity.}
Confining the initial search to Scopus risks missing papers with non-standard terminology. We mitigate this through multiple query iterations and exhaustive forward and backward snowballing, expanding the corpus from 15 to 60 resources. The taxonomy may also include patterns with marginal energy impact; we mitigate this by extracting patterns from peer-reviewed, highly cited sources and established standards. While domain-specific implementations may be absent, specific bad practices typically function as instances of existing subcategories, and the framework is extensible to accommodate new root causes as they emerge.

\textbf{External validity.}
The Pie-Perf dataset focuses on single-threaded, algorithmic competitive programming, explaining the low prevalence of C9 (External Data) and C11 (Concurrency). We mitigate this by deriving the taxonomy from cross-domain literature; the dataset validates 84.6\% of subcategories, with absent entries mapping to missing domain features rather than taxonomic gaps. Empirical validation uses only Python, so prevalence and energy magnitudes reflect CPython behavior. We mitigate this by keeping all definitions language-agnostic, abstracting root causes from Python-specific syntax so the underlying inefficiencies generalize to other high-level languages.

\section{Conclusions and Future Work}
\label{sec:conclusion}

This study addresses the lack of standardization in defining, reporting, and resolving software energy inefficiencies. We conduct a systematic literature review and apply grounded theory coding to develop a language-agnostic, two-level taxonomy of twelve high-level energy smells and 65 actionable root causes. We then empirically validate the framework on profiled Python code pairs using a multi-step LLM classification pipeline. Similar to traditional code smells, our taxonomy establishes a shared vocabulary and practical guidelines for sustainable software development.

Building on this taxonomy and dataset, we plan to validate its cross-language applicability and evaluate smell prevalence across diverse domains. We will also mine real-world open-source repositories (\eg{} GitHub) to measure how often these root causes occur and are refactored in practice. Finally, we aim to develop a lightweight smell detector integrated in popular IDEs for real-time detection and train generative AI models to automatically identify, explain, and refactor energy-inefficient code.

\section*{Acknowledgment}

The authors thank \textit{Kingsley Eneja} for contributions to preliminary explorations related to this work.

\bibliographystyle{IEEEtran} 
\balance
\bibliography{references}

\end{document}